\documentclass[prb,twocolumn,superscriptaddress]{revtex4}
\usepackage{graphicx}
\usepackage{dcolumn}
\usepackage{bm}
\usepackage[latin1]{inputenc}
\usepackage{amsfonts}
\usepackage{amssymb}
\usepackage{amsmath}
\begin{document}
\title{Antiferromagnetic order in (Ga,Mn)N nanocrystals: A density functional theory study}      

\author{C. Echeverr\'ia-Arrondo}
\affiliation{Departamento de F\'isica, Universidad P\'ublica de Navarra, E-31006, Pamplona, Spain}
\affiliation{Donostia International Physics Center (DIPC),  E-20018, San Sebasti\'an/Donostia, Spain}
\author{J. P\'erez-Conde}
\affiliation{Departamento de F\'isica, Universidad P\'ublica de Navarra, E-31006, Pamplona, Spain}
\author{A. Ayuela}
\affiliation{Centro de F\'isica de Materiales CFM-MPC Centro Mixto CSIC-UPV/EHU,  Departamento de
F\'isica de Materiales, E-20018, San Sebasti\'an/Donostia, Spain}
\affiliation{Donostia International Physics Center (DIPC),  E-20018, San Sebasti\'an/Donostia, Spain}

\date{\today}

\begin{abstract}
We investigate the electronic and magnetic properties of (Ga,Mn)N nanocrystals using the density functional theory. We study both wurtzite and zinc-blende structures doped with one or two substitutional Mn impurities. For a single Mn dopant placed close to surface, the behavior of the empty Mn-induced state, hereafter referred to as ``Mn hole'', is different from bulk (Ga,Mn)N. The energy level corresponding to this off-center Mn hole lies within the quantum dot gap near the conduction edge. For two Mn dopants, the most stable magnetic configuration is antiferromagnetic, and this result was unexpected since (Ga,Mn)N bulk shows ferromagnetism in the ground state. The surprising antiferromagnetic alignment of two Mn spins  is ascribed also to the holes linked to the Mn impurities that approach the surface. Unlike (Ga,Mn)N bulk, these Mn holes in confined (Ga,Mn)N nanostructures do not contribute to the ferromagnetic alignment of the two Mn spins.
\end{abstract}

\maketitle




\section{Introduction}


Wide band-gap nitride semiconductors are currently used in full-color displays, white light sources, and ultraviolet laser diodes for high-density storage systems.\cite{orton} Such semiconductors combine group-V nitrogen with elements of group III such as boron, aluminium, gallium, and indium. Recently, the well-known nitride compound GaN has been extensively investigated in the form of quantum dots, both with  wurtzite\cite{widmann,sreekar} and zinc-blende\cite{micic99} crystal structures. The typical phenomena appearing in quantum dots are the discretization of the electronic spectra and the blue shift of the fundamental gaps.\cite{wang42,wang91,albe,albe57,perez110,sapra,carlos-tb} Moreover, the GaN nanocrystals can be doped with diluted magnetic impurities such as manganese. In fact, Ref.~12 shows (Ga,Mn)N quantum dots prepared under solvothermal conditions in the wurtzite phase. These particles seem to show a ferromagnetic signal in the ground state,\cite{biswas} like bulk (Ga,Mn)N as calculated for diluted Mn spins.\cite{das,SIC-2,apl-ayuela} We must note that many experiments on bulk (Ga,Mn)N suggest that Mn spins are not diluted but forming clusters which give the observed ferromagnetism.\cite{f1,f2} A comparative \textit{ab-initio} study between (Ga,Mn)N nanocrystals in wurtzite and zinc-blende phases is nevertheless missing.

In this work we investigate wurtzite and zinc-blende GaN  quantum dots doped with one or two Mn impurities within density functional theory. The doping of nanocrystals with Mn atoms that replace host cations (Mn$_{\rm{Ga}}$) is actually possible, as already confirmed by several experiments on dots larger than ours.\cite{norris,besombes,biswas,hofmann2007} The experimental doping of small nanocrystals of about 1 nm in size has not yet been reported. However, it is expected from evidence of 2 nm undoped nanoparticles already synthesized. Hence, our calculations anticipate future experimental work with small doped dots. The theoretical and computational details are given in Sec.~\ref{computational}. The case of (Ga,Mn)N nanoparticles doped with a single Mn impurity is studied in Sec.~\ref{nanosingle}. We show in this section that the doping reaction is endothermic and requires high temperatures as already confirmed by the experiments.\cite{biswas} The case of (Ga,Mn)N crystallites doped with two Mn dopants is investigated in Sec.~\ref{nanotwo}, where we show that the ground-state Mn impurities are antiferromagnetically aligned. The antiferromagnetic order of the two Mn spins is related to the different chemical environment around the empty state induced by the Mn dopant close to surface. Hereafter, we call ``holes'' these empty Mn-induced states which have mainly Mn $d$ character, but also N.\cite{schneider,ayuela1,ayuela2,marcet} This hole state lies within the nanocrystal gap near the conduction region and therefore does not contribute effectively to the ferromagnetic alignment of the two Mn impurities. The interesting role of the Mn holes near the surface is explained in Sec.~\ref{nanosingle} and Sec.~\ref{nanotwo}. Then, (Ga,Mn)N quantum dots of about 2 nm in size would be likely antiferromagnetic. This possibility is interesting to be taken into account for spintronic applications based on such III-V Mn-doped nanostructures. Of course, the antiferromagnetic behavior of such small dots will also affect the overall magnetic character of granular solids\cite{granular-films, granular-2,granular}  formed by (Ga,Mn)N nanocrystals.

\section{Theory and Computations}
\label{computational}
We calculate (Ga,Mn)N quantum dots within density functional theory, following the Kohn-Sham scheme and the projector augmented-wave method, as implemented in VASP (Vienna Ab-initio Simulation Package).\cite{kresse1,kresse2,kresse3} Apart from the $sp$ valence states inherent in semiconductors, we also take into account the Mn $3d$ states. These latter electrons are responsible for the spin-splitted states at the gap edges through the $sp$-$d$ hybridization. For the exchange-correlation potential\cite{pbe} in the Kohn-Sham equations we use the generalized-gradient approximation $+U$ (GGA+$U$),\cite{u1,SIC-1,u3,u4,ggau,sic} in which $U$ and $J$ are special parameters that account for the strong Coulomb and exchange interactions between the Mn 3$d$ electrons. We take $U=4$ and $J=0.8$ as calculated for Mn when doping bulk GaN.\cite{park} 

We investigate quasi-spherical nanocrystals of $12$~{\AA} in diameter centered in a Ga position both with wurtzite and zinc-blende structures. We passivate the surface dangling bonds with pseudohydrogens (H$^\ast$)\cite{chelikowsky} as an approach to quantum dots synthesized in  colloidal solutions and also grown in semiconductor matrices. These fictitious atoms also prevent the appearance of surface states in the near-gap spectrum.\cite{chelikowsky} Through passivation, every Ga (4$s^2$4$p$) dangling bond at the dot surface is attached to a pseudohydrogen with a fractional charge of $5e/4$, and every N (2$s^2$2$p^3$) dangling bond to a pseudohydrogen with a fractional charge of $3e/4$. Dot surfaces are perfectly saturated and free of defects so that we can avoid any perturbation and clearly investigate Mn-Mn magnetic interactions within nanocrystals. Anyhow, the influence of surface defects in the magnetic properties of Mn-doped dots is an interesting issue for further studies. The doped (Ga,Mn)N nanoparticles contain one or two Mn atoms which substitute for one or two Ga cations.


We use the supercell approximation\cite{scell-1,scell-2,scell-3,scell-4} to calculate wurzite and zinc-blende crystallites which are infinitely repeated in space. The size of the supercell is fixed to 22~{\AA} so that surfaces of adjacent dots become separated by 10~{\AA} and total energies are converged to meVs. This supercell size thus permits an accurate enough description of Mn-Mn magnetic interactions within nanocrystals. The atomic positions are fully relaxed until the forces on the atoms are small enough ($<0.02$~eV/{\AA}). The input Ga-N bond lengths are taken from bulk GaN in the relaxed wurtzite and zinc-blende structures, $d_{Ga-N}=1.99$~{\AA}. For wurtzite GaN, our calculated lattice constants are $a$=3.24~{\AA} and $c$=5.29~{\AA}; for zinc-blende GaN, our lattice constant is $a$=4.59~{\AA}. The cut-off energy in the plane-wave basis set is fixed to 500 eV in order to converge the total energies of bulk GaN and ferromagnetic MnN below 1 meV.

\begin{figure}[t!]
\includegraphics[scale=.8]{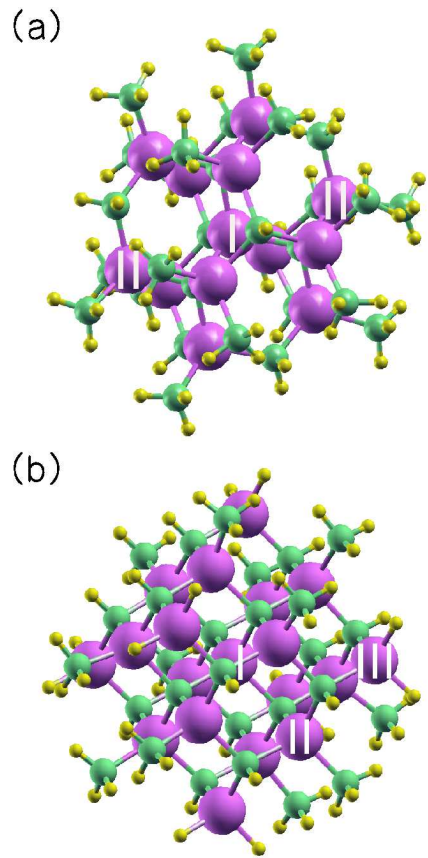}
\caption{\label{fig:fig1}(Color online)  Studied (Ga,Mn)N nanocrystals with (a) wurtzite and (b) zinc-blende structures. The Ga atoms are drawn with large spheres (pink), the N atoms with medium spheres (green),  and the  $\rm{H}^{\star}$ pseudohydrogens with small  spheres (yellow). The  considered positions for the substitutional Mn impurities in Ga sites are the centers labeled ``I'', the off-center sites labeled ``II'', and for zinc-blende dots the sites labeled ``III''.}
\end{figure}

\section{Nanoparticles with a single Mn impurity}
\label{nanosingle}
In this section we study GaN nanocrystals doped with a single Mn atom. The dopant concentrations\cite{fatah} are $x=1/13\simeq 0.08$ for wurtzite quantum dots and $x=1/19\simeq 0.05$ for zinc-blende dots. These concentrations are in the same order of those used in the  experiments.\cite{biswas} The relaxed geometries are depicted in Fig.~\ref{fig:fig1}, where we see that the Mn impurity can occupy the central site labeled ``I'', the off-center sites labeled ``II'', and in zinc-blende particles also the sites labeled ``III''. Positions II are near the crystal surface, not exactly at the surface. Since results concerning site III are dependent on the passivation species, in the following discussions we will focus on sites I and II.

\subsection{Geometric expansion of N atoms around the Mn impurity}

As we relax the atomic positions in (Ga,Mn)N quantum dots, we comment the influence of Mn doping on the crystal geometries. In the undoped nanoparticles the Ga-N bond lengths calculated for wurtzite and zinc-blende structures are $2.02-2.04$~{\AA} for the central Ga cations and $1.96-2.00$~{\AA} around the off-center Ga ions placed in site II. The smaller bond distances for position II as compared with position I are typically due to quantum dot surfaces. In the doped nanoparticles, the Mn-N bonds  measure $2.06-2.08$~{\AA} for the central-Mn case and $1.99-2.04$~{\AA} around the off-center Mn dopants placed in site II. Therefore, as compared with the undoped structures, the N shell around Mn expands by $2-3\%$, in close accordance with the 2$\%$ calculated expansion in wurtzite and zinc-blende bulk (Ga,Mn)N. We note that this expansion was not observed in II-VI (Cd,Mn)Te quantum dots,\cite{carlos} and is even contrary to the contraction around Mn dopants in III-V (In,Mn)P nanowires.\cite{nanowire}

\subsection{Quantum dot stability versus Mn position}
\label{stability}

We focus now on the energies involved in the formation and doping of (Ga,Mn)N nanocrystals. The cohesive energy of the undoped nanoparticle is defined in relation to the free atoms, $-(E_0-\sum_i^N E_{\rm{at}}^i)/N$, where $E_{\rm{at}}^i$ is the energy of the $i$th free atom, $E_0$ is the total ground-state energy, and $N$ is the number of atoms in the dot. Single-atom total energies are also obtained within the supercell approximation. The computed values are 3.607~eV/atom for the wurtzite structure and 3.593~eV/atom for the zinc-blende one. In undoped nanocrystals the wurtzite phase is thus more stable than the zinc-blende phase. However, since the growth of quantum dots (QDs) also depends on kinetics and other chemical potentials apart from those used in the cohesive energy, the undoped GaN crystallites can actually be synthesized both with wurtzite and zinc-blende structures.\cite{widmann,sreekar,micic99} Therefore, we are going to study Mn-doped GaN nanoparticles in both kind of geometries, wurtzite and zinc blende.

\begin{figure}[t!]
\includegraphics[scale=.5]{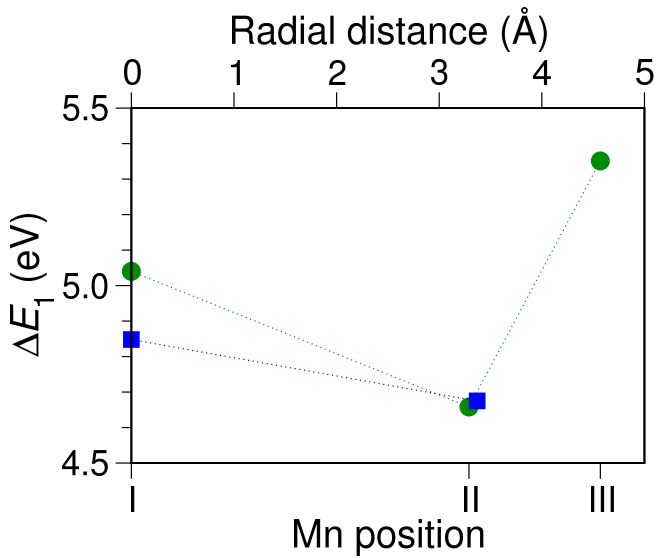} 
\caption{\label{fig:fig5}(Color online) Substitutional reaction energy for a single Mn impurity. Squares (blue) indicate wurtzite structure and  circles (green) indicate zinc-blende structure. Dotted lines are drawn to guide the eye. The most stable position for the Mn dopant is site II irrespective of the crystal geometry.}
\end{figure}

Anyhow, the energetics of Mn doping for both structures is studied by the following reaction,
\begin{equation}
\label{ener}
\textrm{GaN}~\textrm{QD}+\textrm{Mn}^{+2}\longrightarrow\textrm{(Ga,Mn)N}~\textrm{QD}+\textrm{Ga}^{+2},
\end{equation}
in which the Mn$^{+2}$ dopant substitutes a Ga$^{+3}$ cation in the quantum dot. Reaction (1) points to the main energy difference in the process of solvothermal growth during substitutional doping. The cationic form for the Mn dopant participates in reactions with the solvent which are beyond the scope of this work. For instance, these reactions may include precursors such as GaCl3 and MnCl2 in the presence of hexamethyldisilazane (HDMS).\cite{biswas} Total energies of ions stem from the calculation of total energies of neutral atoms to which the first and second ionization energies (IEs) are summed up, that is, $E_0$(Mn$^{+2}$)$=E_0$(Mn)+1$^{\rm{st}}$~IE + 2$^{\rm{nd}}$~IE, where IE values are taken from the literature. The substitutional energy $\Delta E_1$ involved in reaction (\ref{ener}) is calculated against the Mn position and plotted in Fig.~\ref{fig:fig5}.  Since this energy is positive, reaction (\ref{ener}) becomes endothermic and its activation requires high temperatures. This finding is in agreement with the experiments in which (Ga,Mn)N nanocrystals are prepared under solvothermal conditions at about 350 $^{\circ}$C.\cite{biswas} In addition, the substitutional energy $\Delta E_1$ is smaller for site II than for site I. This indicates that position II near the surface is more stable than position I, in accordance with previous calculations for substitutional Mn impurities embedded in II-VI QDs\cite{carlos} and also in III-V nanowires.\cite{nanowire}

\begin{figure}[t!]
\includegraphics[scale=1]{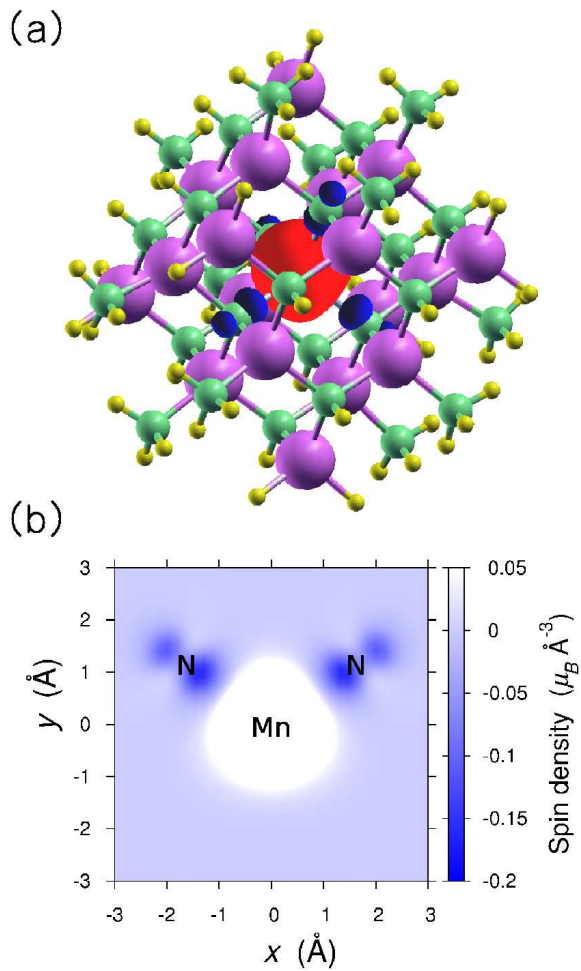}
\caption{\label{fig:fig2}(Color  online) (a) Spin density in zinc-blende quantum dots doped with a central Mn impurity. The isosurfaces correspond to constant densities $\pm0.05\mu_{B}${\AA}$^{-3}$, with light gray (red) for the positive value and dark gray (blue) for the negative one. (b) The shown spin density is in a plane which includes the Mn dopant and two of its four N  neighbors. The density at the Mn site is chopped for the sake of clarity. See how the spherical integration of the spin density around Mn and N atoms yields local magnetic moments antiferromagnetically coupled.}
\end{figure}

\subsection{Local magnetic moments at the Mn and neighbor N sites}

When doping (Ga,Mn)N nanocrystals with a single Mn impurity, one $3d$ electron of the free Mn atom ($3d^5$) transfers to its neighbor N anions and yields a $3d^4$ configuration at the dopant position with a localized Mn hole.\cite{SIC-1,marcet} Nevertheless, the value of the local magnetic moment at the Mn site is different from 4$\mu_{B}$ due to the $sp$-$d$ hybridization. This local moment is 3.99$\mu_{B}$/4.10$\mu_{B}$ when the Mn atom occupies the center of wurtzite/zinc-blende quantum dots and 3.88$\mu_{B}$/3.88$\mu_{B}$ when it is placed off-center in position II. Moreover, the Mn dopant induces the magnetic polarization of its neighbor N anions. In Fig.~\ref{fig:fig2} we show for zinc-blende nanoparticles the spin density around the central Mn impurity and its four N neighbors. The integration of this density within spheres of Wigner-Seitz radii centered in the N atoms results in N local magnetic moments of -0.06$\mu_{B}$/-0.09$\mu_{B}$ for wurtzite/zinc-blende quantum dots. Small modifications of these radii would lead to roughly the same local moments with the same signs, as already seen for other Mn-doped nanocrystals.\cite{carlos} The exchange coupling between Mn and N magnetic moments is thus antiferromagnetic, as also calculated for bulk wurtzite and bulk zinc-blende (Ga,Mn)N.

We can compare the Mn local magnetic moments obtained for quantum dots and for bulk (Ga,Mn)N with similar Mn concentration. For bulk wurtzite $\rm{Ga}_{0.917}\rm{Mn}_{0.083}N$, the computed Mn magnetic moment is 3.99$\mu_{B}$; for bulk zinc-blende $\rm{Ga}_{0.937}\rm{Mn}_{0.063}N$, the Mn magnetic moment is 4.03$\mu_{B}$. These bulk values are hence similar to the previous Mn magnetic moments centered in wurtzite and zinc-blende (Ga,Mn)N nanocrystals. However, they show larger differences with respect to those moments calculated for Mn in site II near the surface. We shall see in next Sec.~\ref{gap-} the local densities of states for further explanation about such larger differences.

\begin{figure}[t!]
\includegraphics[scale=1.0]{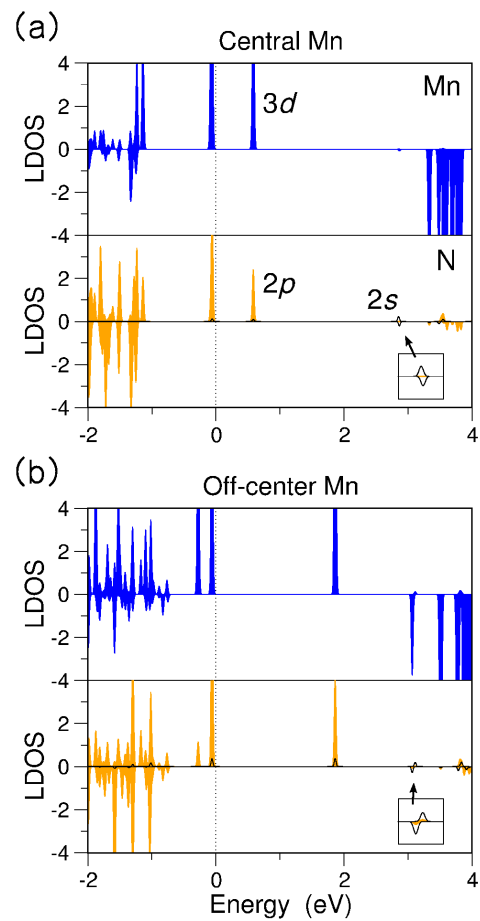}
\caption{\label{fig:fig3}(Color online) Local densities of states (LDOS) for wurtzite quantum dots doped with a single Mn impurity placed (a) in the dot center and (b) in site II off-center. The densities of states are projected onto the Mn $3d$ states in black (blue), the neighbor N $2s$ states with solid black lines, and the neighbor N $2p$ states in gray (orange). The vertical dotted lines indicate the Fermi energy. See for Mn off-center the hole level placed within the gap at around 2~eV near the conduction edge. The zinc-blende densities of states are analogous to the wurtzite ones with the only difference of a triply degeneracy in the spin-up valence edge state (HOMO) for the central-Mn case. Insets widen the conduction edge states (ULUMOs). Note the Mn hole level closer to the conduction region in panel (b) as compared with panel (a), and the different energy shifts for the up and down ULUMOs.}
\end{figure}

\subsection{Role of Mn hole in nanostructures}
\label{gap-}

We investigate the hole linked to the Mn impurity by looking at the local densities of states for Mn in site I and site II close to surface. In Fig.~\ref{fig:fig3} we show the local projections of the crystal states onto the  Mn $3d$ states and N $sp$ states around the Mn dopant. The wurtzite case is given in Fig.~\ref{fig:fig3}(a) for a centered Mn impurity and in Fig.~\ref{fig:fig3}(b) for an off-center Mn placed in site II. From the densities of states we conclude that (i) the hole level associated with the central Mn dopant lies at around 0.6~eV near the Fermi energy; (ii) nevertheless, the hole level associated with the off-center Mn impurity lies at around 2~eV near the conduction region. 

\begin{figure}[t!]
\includegraphics[scale=1.0]{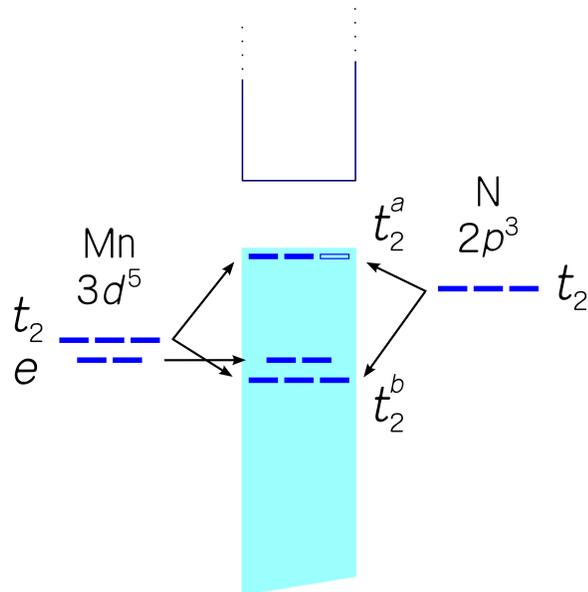} 
\caption{\label{fig:fig4}(Color online) Branch diagram describing the $sp$-$d$ hybridization in a zinc-blende quantum dot doped with a centered Mn impurity. On the left we show the splitted Mn $3d$ states in the $T_d$-symmetry crystal field, in the middle the hybridized $pd$ bonding and antibonding nanocrystal states, and on the right the spin-up 2p states of the four N atoms surrounding the Mn impurity in the Td-symmetry crystal field. The Mn hole is indicated as an open rectangle in the shadowed valence region.}
\end{figure}

The zinc-blende densities of states are similar to the wurtzite densities apart from the central-Mn case which shows a triply degeneracy in the spin-up valence edge state. Such zinc-blende case is described in detail in Fig.~\ref{fig:fig4}. Due to the $T_d$ crystal field, the five 3$d$ states of the central Mn atom are divided in two groups; one is composed of three $t_2$-symmetry states and the other of two $e$-symmetry states. The $t_2$ states of the Mn impurity hybridize with the $p$-like $t_2$ states of its neighbor N atoms and yield the formation of bonding and antibonding states. The latter states are degenerate and only partially occupied due to the Mn hole. For zinc-blende nanoparticles with off-center Mn dopants and also for wurtzite dots, the crystal symmetry is $T_d$-like.

\subsubsection{Gap-edge splittings and related $sp$-$d$ exchange constants}
\label{gap-split}
The $sp$-$d$ hybridization between the Mn 3$d$ states and the $sp$ host states yields an effective Mn-quantum dot exchange interaction that splits the crystal states at the gap edges.\cite{larson} Note that we are interested in the change of III-V dot states by Mn impurities. The $ab$-$initio$ energy splittings are explicitly given here and also rewritten in terms of $sp$-$d$ constants, $N_0\alpha$ and $N_0\beta$. Thereby, the rescaled splittings are interesting not only for theoreticians but also for experimentalists, both working on diluted magnetic compounds in the bulk\cite{gaj,larson} and in quantum dots.\cite{bhatta,gamelin,santangelo}

\begin{figure}[t!]
\includegraphics[scale=.4]{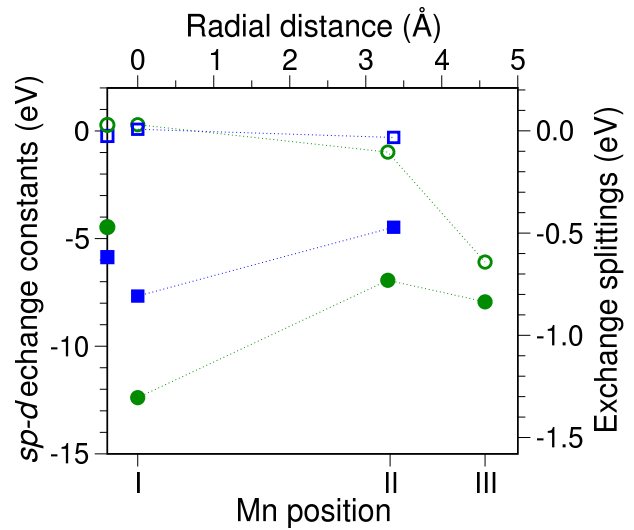}
\caption{\label{fig:fig7}(Color online) The $sp$-$d$ exchange constants and gap-edge splittings for single Mn-doped nanocrystals as a function of the impurity position. The open symbols stand for $N_0\alpha$ exchange constant and the closed symbols stand for $N_0\beta$. Squares (blue) refer to wurtzite structure and circles (green) refer to zinc-blende structure. Dotted lines are drawn to guide the eye. The bulk exchange constants are indicated on the left axis with similar notation. The $|N_0\beta|$ values for Mn in the dot center (site I) are larger than the bulk ones due to the different roles of the Mn holes.}
\end{figure}

The $sp$-$d$ exchange constants are defined with the following mean-field theory expressions:\cite{bhatta83,larson,gamelin,merad,ayuela1,ayuela2}

\begin{equation}
\label{eq:exchange}
N_0\alpha=\frac{\bigtriangleup E^c}{x\langle  S_z\rangle}\;\;\textrm{and}\;\;N_0\beta=\frac{\bigtriangleup E^v}{x\langle S_z\rangle}.
\end{equation}
Here $N_0$ is the number of cations per unit volume;\cite{bhatta3} $\bigtriangleup E^{c}=E^{c}\textrm{(spin down)}-E^{c}\textrm{(spin up)}$ is the splitting of the up and down Undoped Lowest Unoccupied Molecular Orbitals (ULUMOs); $\bigtriangleup E^{v}=E^{v}\textrm{(spin down)}-E^{v}\textrm{(spin up)}$ is the splitting of the up and down Highest Occupied Molecular Orbitals (HOMOs); and $\langle  S_z\rangle=4/2$ is the average Mn spin. The splittings $\bigtriangleup E^{c}$ and $\bigtriangleup E^{v}$ can be extracted from the local densities of states depicted in Fig.~\ref{fig:fig3}. The $N_0\alpha$ and $N_0\beta$ exchange constants are calculated from Eq.~(\ref{eq:exchange}) and presented in Fig.~\ref{fig:fig7} together with $\bigtriangleup E^{c,v}$ for different Mn positions. Both wurtzite and zinc-blende $|N_0\beta|$ values are larger for the central site I than for site II off-center, as it also happens for II-VI quantum dots doped with Mn.\cite{carlos}

\begin{figure}
\includegraphics[scale=.9]{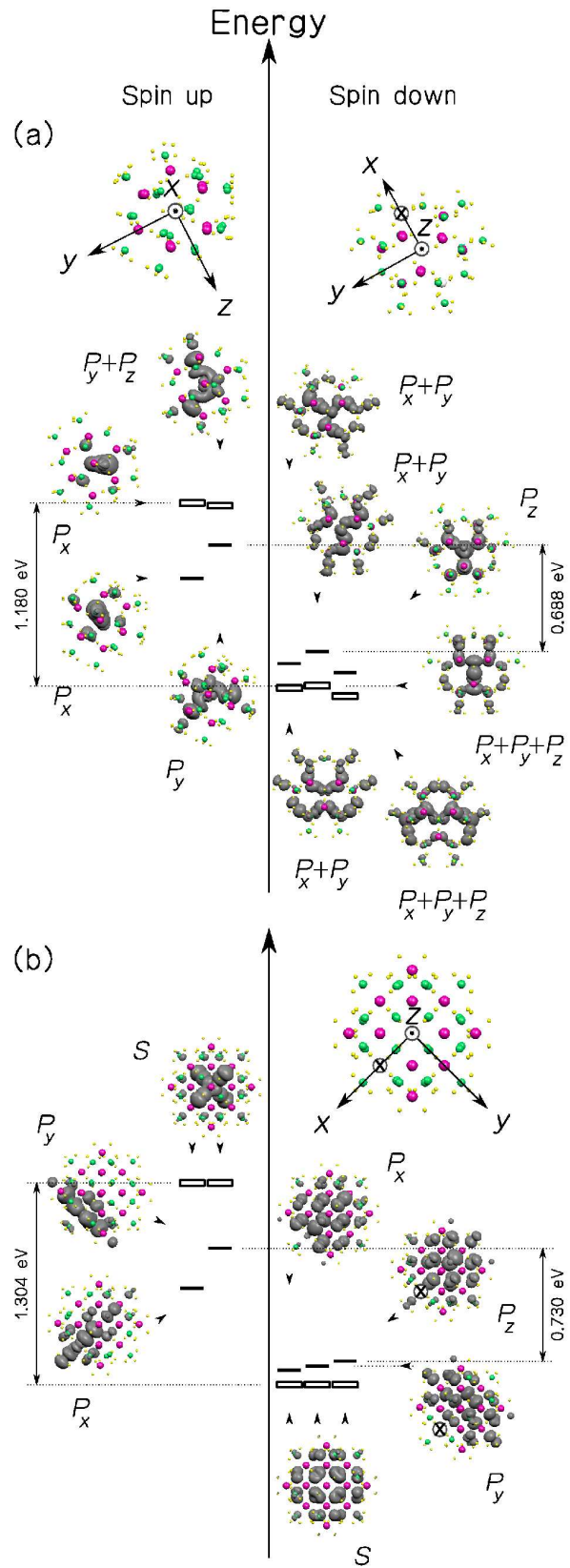}
\caption{\label{fig:fig10}(Color online) Valence levels in the gap region for (a) wurtzite and (b) zinc-blende quantum dots doped with a single Mn impurity located in the central site I with open rectangles and in an off-center site II with solid lines. The charge densities next to the valence states show their global $S$- or $P$-like character. The geometries with the reference axis centered in the dot are oriented so that position II lies on the $x$ axis. Crosses indicate site II whenever visible.} 
\end{figure}

The dependence of the $N_0\beta$ exchange constant with the Mn position can be explained in more detail by looking at the valence states in the gap region. In Fig.~\ref{fig:fig10} we show these states for two Mn sites, dot center and off-center position II. For the central-Mn case, the up-down HOMO splitting is large both for wurtzite and zinc-blende geometries. However, when Mn is moved off-center the spin-up valence levels decrease in energy and the  spin-down levels slightly increase. These shifts are due to a smaller Mn-quantum dot exchange interaction caused by a smaller host charge density around position II. The up-down HOMO splittings for these Mn sites are consequently reduced as compared with those for the dot centers, and the $N_0\beta$ exchange constants become hence smaller. We note that for the central-Mn case the charge of the up states is mainly distributed over the Mn impurity and its N neighbors, but for the down states it suffers a deplection around the Mn dopant that we ascribe to the fact that Mn and N are antiferromagnetically coupled. In addition, for Mn in position II the charge in the up HOMOs spreads perpendicularly to the $x$ axis and globally shows $P_y$-like character.

The $sp$-$d$ exchange constants are different for nanoparticles than for bulk (Ga,Mn)N. In order to assess this difference, we calculate $N_0\alpha$ and $N_0\beta$ for two bulk compounds with Mn concentrations which are similar to those in the studied crystallites. For bulk wurtzite $\rm{Ga}_{0.917}\rm{Mn}_{0.083}N$ we obtain $N_0\alpha=-0.23$~eV and $N_0\beta=-5.86$~eV; for bulk zinc blende $\rm{Ga}_{0.937}\rm{Mn}_{0.063}N$ we obtain $N_0\alpha=0.28$~eV and $N_0\beta=-4.47$~eV. For comparison, these bulk constants are indicated in Fig.~\ref{fig:fig7} as marks on the left axis. The $N_0\alpha$ values for Mn in positions I and II are comparable to the $N_0\alpha$ ones calculated for bulk wurtzite and bulk zinc blende. Moreover,  the $|N_0\beta|$ values are larger than the bulk ones for Mn in the central site I than in site II near the surface. Due to confinement, the hole level associated with the central Mn impurity is closer to the valence levels within the nanocrystal gap than in the bulk. The stronger exchange interaction between this Mn hole with $d$-like character and the valence edge states increases the splitting of the up and down HOMOs and consequently the $|N_0\beta|$ values. For position II as compared with position I, the off-center Mn hole lying near the conduction levels indicates a smaller interaction with the valence states, and hence smaller $|N_0\beta|$ values.


\section{Nanocrystals with two Mn impurities. Antiferromagnetic order in the ground state}
\label{nanotwo}

In this section we investigate (Ga,Mn)N nanoparticles doped with two substitutional Mn impurities in the ferromagnetic and antiferromagnetic configurations. The ferromagnetic state is calculated for a total magnetic moment in the quantum dot (QD) of 8$\mu_{B}$; the antiferromagnetic state is calculated for a null magnetic moment in the QD. 

\subsection{Nanocrystal stability versus positions and magnetic alignments of the two Mn spins}

\begin{figure}[t!]
\includegraphics[scale=.4]{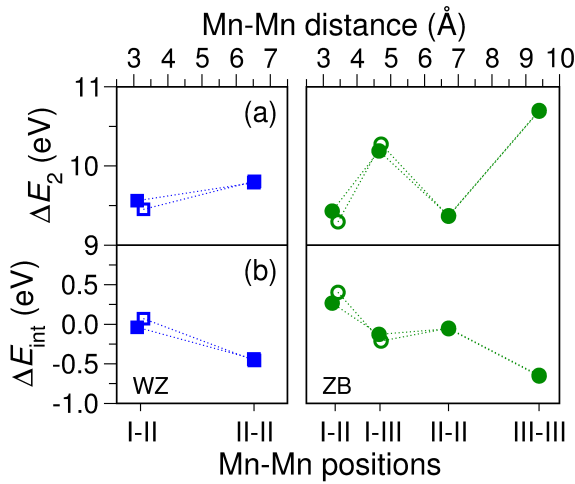}
\caption{\label{fig:fig6}(Color online) (a) Substitutional reaction energy for two Mn dopants. (b) Interaction energy between two Mn impurities. Squares (blue) on the left stand for wurtzite (WZ) structure and circles (green) on the right stand for zinc-blende (ZB) structure. Closed symbols refer to ferromagnetic Mn spins and open symbols refer to antiferromagnetic. Dotted lines are drawn to guide the eye. We note from (a) that the most stable Mn impurities are aligned antiferromagnetically and located in positions I-II both for  wurtzite and zinc-blende geometries.}
\end{figure}

By doping with two Mn ions we replace two Ga cations as described by the following reaction: 

\begin{equation}
\textrm{GaN}~\textrm{QD}+2\textrm{Mn}^{+2}\longrightarrow\textrm{(Ga,Mn)N}~\textrm{QD}+2\textrm{Ga}^{+2}.
\end{equation} 
The required energy for this double substitution is referred to as $\Delta  E_2$ and plotted in Fig.~\ref{fig:fig6}(a) for different Mn-Mn positions and magnetic alignments. As commented previously, the positive substitutional energies indicate that reaction (3) is endothermic and thus activated by increasing the temperature, as it occurs in the experiments.\cite{biswas} Fig.~\ref{fig:fig6} also shows that the most stable Mn impurities are aligned antiferromagnetically and placed in sites I-II close to  surface. The calculated antiferromagnetic ground state was unexpected, since it is different from the ferromagnetic alignment of Mn spins in bulk (Ga,Mn)N.\cite{das,SIC-2,apl-ayuela} We relate it to the different role of the hole linked to the Mn dopant that approaches the surface. The energy of this hole lies within the nanocrystal gap near the conduction levels. Therefore, this Mn hole does not contribute effectively to the ferromagnetic order of the two Mn spins. As a consequence, the Mn-Mn coupling becomes antiferromagnetic in the ground state as it occurs for two Mn spins in II-VI (Cd,Mn)Te QDs.\cite{carlos}

We next study the interaction energy between the two Mn dopants which is defined as 

\begin{equation}
\Delta E_{\rm{int}}=\Delta  E_1(\textrm{Mn}_1)+\Delta  E_1(\textrm{Mn}_2)-\Delta  E_2;
\end{equation}

\noindent Mn$_1$ stands for the first Mn atom and Mn$_2$ stands for the second Mn atom. The calculated interaction energies are plotted in Fig.~\ref{fig:fig6}(b) as a function of the positions and magnetic couplings of the two Mn spins. The  interaction energy $\Delta E_{\rm{int}}$ quantifies the relative stability of nanoparticles doped with one or two Mn impurities. Positive interaction values indicate that the two Mn dopants tend to occupy the same nanocrystal and negative values indicate that they tend to dope two different quantum dots individually. For instance, from Fig.~\ref{fig:fig6}(b) it can be seen that one crystallite with two antiferromagnetic Mn spins in positions I-II is more stable than two nanoparticles with two single Mn spins placed in sites I and II. We stress again that in doping reactions kinetics and other chemical potentials different from those used in the cohesive energy could also play an important role and may modify previous results concerning the stability of the nanostructures. 

\begin{figure}[t!]
\includegraphics[scale=.4]{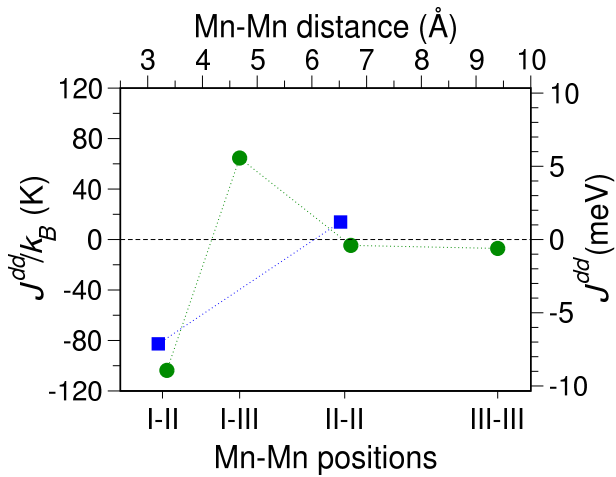}
\caption{\label{fig:fig9}(Color online) The $J^{dd}$ exchange constants for two Mn impurities as a function of the Mn-Mn positions. Squares (blue) denote wurtzite structure and circles (green) denote zinc-blende structure. Dotted lines are drawn to guide the eye. For the most stable positions I-II the negative $J^{dd}$ values indicate that the two Mn spins are unexpectedly ordered antiferromagnetically, unlike Mn spins in bulk ferromagnetic (Ga,Mn)N.}
\end{figure}

Since we are dealing with two Mn impurities, we calculate their local magnetic moments and total energies in the ferromagnetic and antiferromagnetic states.  They are not explicitely given here but rewritten in terms of an effective Mn-Mn exchange interaction, quantified by $J^{dd}$, which is interesting not only for theoreticians\cite{larson,chanier,carlos} but also for experimentalists\cite{shapira,stowell,sati,white} working on diluted magnetic semiconductors. The $J^{dd}$ exchange constant stems from the Heisenberg-like Hamiltonian\cite{larson} $H=-2J^{dd}\rm{\textbf{S}_1}\rm{\textbf{S}_2}$ and it is thus defined as 
\begin{equation}
J^{dd}=-\frac{       2 (E_0^{\rm{FM}}  -  E_0^{\rm{AFM}})     }{    \mu_{\rm{Mn_1}}^{\rm{FM}}   \mu_{\rm{Mn_2}}^{\rm{FM}}+\mu_{\rm{Mn_1}}^{\rm{AFM}}\mu_{\rm{Mn_2}}^{\rm{AFM}}   },
\end{equation} 
where $E_0^{\textrm{FM}}$ is the total energy of the ferromagnetic state, $E_0^{\textrm{AFM}}$ is the total energy of the antiferromagnetic state, and $\mu_{\rm{Mn_{1}}}=2S_{1}$  is the local magnetic moment at the Mn$_{1}$ site in Bohr magnetons. The $J^{dd}$ exchange constants are calculated and plotted in Fig.~\ref{fig:fig9} as a function of the Mn-Mn positions. Negative $J^{dd}$ values mean antiferromagnetic alignments between Mn spins and positive $J^{dd}$ values, ferromagnetic alignments. Fig.~\ref{fig:fig9} shows also that in the most stable positions I-II the two Mn atoms are antiferromagnetically ordered unlike Mn spins in bulk ferromagnetic (Ga,Mn)N.

\begin{figure}[t!]
\includegraphics[scale=.45]{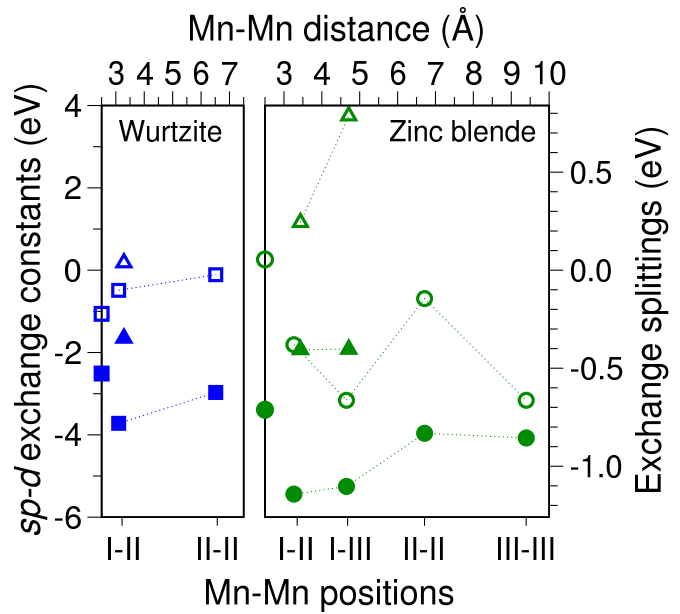}
\caption{\label{fig:fig8}(Color  online) The $sp$-$d$ exchange constants and gap-edge splittings for quantum dots doped with two Mn impurities. Squares (blue) on the left refer to wurtzite quantum dots in the ferromagnetic configuration of Mn spins and circles (green) on the right refer to zinc-blende dots. Open symbols denote $N_0\alpha$ exchange constant and closed symbols denote $N_0\beta$. The triangles stand for antiferromagnetic Mn spins. Dotted lines are drawn to guide the eye. For comparison, the exchange constants for bulk ferromagnetic (Ga,Mn)N are indicated on the two left axis with similar notation. The $|N_0\beta|$ values are larger for ferromagnetic nanocrystals than for bulk (Ga,Mn)N.}
\end{figure}

\subsection{The $sp$-$d$ exchange constants and Mn holes}

To look at the modification of the III-V dot states by Mn impurities, we now investigate the $N_0\alpha$ and $N_0\beta$ exchange constants obtained from the calculated spin splittings at the gap edges and Eq.~(\ref{eq:exchange}). The computed values are shown in Fig.~\ref{fig:fig8} as a function of the positions and magnetic couplings of the two Mn spins. The main results are the following: 

(i) $|N_0\alpha|$ and $|N_0\beta|$ values are larger for ferromagnetic Mn spins in positions I-II than for farther apart Mn spins in positions II-II. This decrease is similar to that observed for two Mn spins in II-VI (Cd,Mn)Te quantum dots.\cite{carlos}

(ii) $N_0\alpha$ values are negative for ferromagnetic Mn spins.  The hole levels associated with the off-center Mn impurities lie at high energies within the nanocrystals gaps and push the spin-up ULUMOs above the spin-down ULUMOs.

(iii) Positive $N_0\alpha$ values are larger for antiferromagnetic Mn spins placed in sites I-II than $N_0\alpha$ for a doped nanoparticle with a central Mn atom. The hole level linked to the spin-down Mn dopant located in site II pushes upward in energy the spin-down ULUMO and thereby increases the spin splitting of the ULUMOs corresponding to the single-Mn case.

(iv) Comparing Fig.~\ref{fig:fig7} and Fig.~\ref{fig:fig8} we see that the largest $|N_0\beta|$ values are for wurtzite and zinc-blende quantum dots doped with a central Mn impurity. 

As in the previous discussion of Sec.~\ref{gap-}, we now compare the $N_0\alpha$ and $N_0\beta$ exchange constants for nanocrystals and for bulk (Ga,Mn)N. We calculate the bulk ferromagnetic compounds wurtzite $\rm{Ga}_{0.833}\rm{Mn}_{0.167}N$ and zinc-blende $\rm{Ga}_{0.875}\rm{Mn}_{0.125}N$. The Mn concentration in bulk $\rm{Ga}_{0.833}\rm{Mn}_{0.167}N$ is similar to $x\simeq 0.15$ in wurtzite quantum dots; that of bulk $\rm{Ga}_{0.875}\rm{Mn}_{0.125}N$ is similar to $x\simeq 0.11$ in zinc-blende dots. For bulk wurtzite we obtain $N_0\alpha=-1.06$~eV and $N_0\beta=-2.51$~eV; for bulk zinc blende we obtain $N_0\alpha=0.26$~eV and $N_0\beta=-3.39$~eV. These exchange values are shown for comparison in Fig.~\ref{fig:fig8} as marks on the left axis. For wurtzite nanoparticles in the ferromagnetic configuration of Mn spins, the $|N_0\alpha|$ values are smaller than in the bulk; on the contrary, for zinc-blende quantum dots in the ferromagnetic state, the $|N_0\alpha|$ values are significantly larger than in the bulk. For ferromagnetic nanocrystals, the wurtzite and zinc-blende $|N_0\beta|$ values are both larger than the corresponding $|N_0\beta|$ bulk constants. These differences can also be explained by the different roles played by the Mn holes in bulk structures and in quantum dots (as seen in previous Sec.~\ref{gap-}). It seems that for nanostructures, the situation of the dopant hole within the gap must be analyzed in detail in order to understand their basic magnetic properties.

\begin{figure}[t!]
\includegraphics[scale=.5]{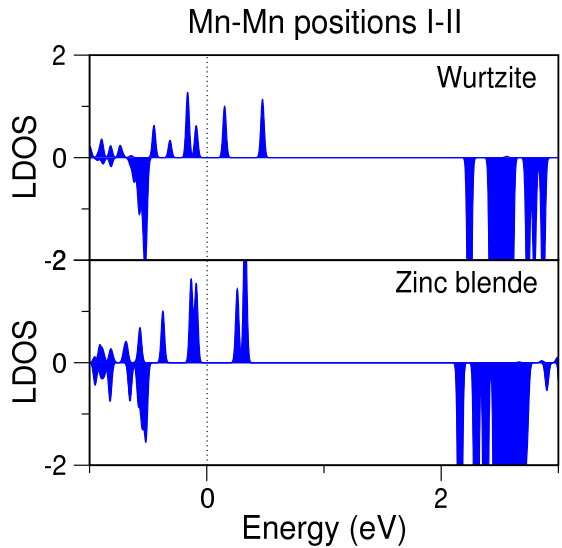}
\caption{\label{fig:fig11}(Color  online) Local densities of states (LDOS) projected onto the Mn 3$d$ states for wurtzite and zinc-blende quantum dots made of (Ga,Mn)As. The two Mn impurities are closely placed in positions I-II and ferromagnetically aligned. Note that the two Mn holes approach each other and the valence edge unlike in (Ga,Mn)N nanoparticles.}
\end{figure}

\section{Comparison with Mn-doped GaAs nanocrystals}

So far, the antiferromagnetic alignment of Mn in GaN quantum dots has been the main result of our discussion: Role of Mn hole in GaN nanostructures in Sec.~\ref{gap-} and magnetic order of two Mn spins in the last section. In fact, the Mn atom close to the crystal surface induces impurity states near the conduction levels. Now, to make contact with the antiferromagnetic coupling of such Mn atoms, it is essential to consider other III-V nanoparticles.

Experimentally, (Ga,Mn)As nanocrystals can be created by Mn implantation on GaAs followed by thermal treatment,\cite{couto} and also by annealing (Ga,Mn)As thin films grown by molecular beam epitaxy.\cite{janik} The self-organized nanoclusters are analyzed by x-ray spectroscopy,\cite{couto} microscopic techniques,\cite{couto,wasik} and also by using SQUID magnetometry.\cite{wasik} Magnetic force microscopy measurements on (Ga,Mn)As nanoprecipitates show ferromagnetic features at room temperature.\cite{wasik} This experimental result and the reported antiferromagnetic behavior of (Ga,Mn)N quantum dots motivates us to search for a change in the magnetic order also in (Ga,Mn)As nanocrystals.

To illustrate this other compound, (Ga,Mn)As, without taking into consideration all the positions for Mn, let us focus on I-II sites, i.e. the case shown to be clearly antiferromagnetic (AFM) in (Ga,Mn)N. We calculate (Ga,Mn)As nanoparticles with two Mn dopants which replace two Ga atoms in the close interacting sites I-II. We study both wurtzite and zinc-blende geometries of about 1~nm in diameter as those given in Fig.~\ref{fig:fig1}. The computational details in this case are similar to those already explained in Sec.~\ref{computational}. 

As compared with the undoped structures, in the doped ones the As shell around the Mn atom close to surface is expanded by $2-4\%$, thus it is more expanded than the N shell in (Ga,Mn)N nanoparticles. Moreover, the computed quantum dots are ferromagnetic (FM) in the ground state, with a FM-AFM exchange interaction of 122~meV in the wurtzite phase and 104~meV in the zinc-blende phase. These exchange energies are roughly half the bulk value for Mn atoms sitting in close Ga positions, $\sim$200~meV,\cite{ayuela1} in agreement with the same decreasing tendency already calculated for other magnetic nanostructures such as (Cd,Mn)Te nanocrystals.\cite{carlos}

It should be clear from the previous discussion of Mn in GaN quantum dots that for ferromagnetic Mn spins the so-called Mn hole levels must be close to the GaAs valence states. Indeed, the local densities of states projected onto the Mn 3$d$ states in Fig.~\ref{fig:fig11} show that the Mn holes in GaAs nanoparticles are close together in the nearby of the valence region. These Mn holes behave hence as in bulk (Ga,Mn)As\cite{SIC-2} and mediate the ferromagnetic alignment of the two Mn spins.

Looking at the Mn hole levels in its relation to the valence (conduction) region, we can see a ferromagnetic (antiferromagnetic) behavior of Mn impurities in doped nanostructures. The difference in this hole position must be found in the smaller bond compression around Mn for (Ga,Mn)As quantum dots.


\section{Conclusions}
\label{conclusion}

In summary, we have investigated wurtzite and zinc-blende (Ga,Mn)N quantum dots doped with one or two substitutional Mn impurities within density functional theory. We have obtained that wurtzite and zinc-blende structures show similar results as they cannot be distinguished up to second neighbors. Anyhow, there are small differences between them commented in the text when appropiate. For a single Mn dopant in the dot center, the calculated $|N_0\beta|$ values are larger than in bulk (Ga,Mn)N with similar Mn concentration. For two Mn dopants, the most stable magnetic state is antiferromagnetic, and this was unexpected since bulk (Ga,Mn)N exhibits ferromagnetism in the ground state. We ascribe this surprising effect in (Ga,Mn)N nanoparticles to the holes linked to the Mn impurities placed close to surface. These holes do not contribute effectively to the ferromagnetic order of the two Mn spins. We show that the ferromagnetic behavior of bulk (Ga,Mn)N can be changed by reducing the crystal size.

From the antiferromagnetic result on small (Ga,Mn)N nanoparticles, it seems possible that larger dots and other nanostructures such as thin films with Mn dopants close to surface could also be antiferromagnetic in the ground state. Therefore, we hope that our results concerning antiferromagnetic (Ga,Mn)N quantum dots will encourage further \textit{ab-initio} calculations and experiments on Mn impurities buried near surfaces of semiconductor nanostructures. Indeed, recent in-progress work on (Ga,Mn)N nanolayers suggests the same antiferromagnetic behavior for Mn spins.

\acknowledgments{This  work  was  supported  by  the  Basque  Government through the NANOMATERIALS
project (Grant No. IE05-151) under the ETORTEK Program (iNanogune), the Spanish Ministerio de
Ciencia y Tecnolog\'ia of Spain (Grant Nos. TEC2007-68065-C03-03 and Fis2007-66711-C02-02, and MONACEM
project), and the University of the Basque Country (Grant No. IT-366-07). The computing resources from  the Donostia  International Physics Center and the SGI-SGIKerUPV are
gratefully acknowledged.}



\begin{thebibliography}{49}
\expandafter\ifx\csname natexlab\endcsname\relax\def\natexlab#1{#1}\fi
\expandafter\ifx\csname bibnamefont\endcsname\relax
  \def\bibnamefont#1{#1}\fi
\expandafter\ifx\csname bibfnamefont\endcsname\relax
  \def\bibfnamefont#1{#1}\fi
\expandafter\ifx\csname citenamefont\endcsname\relax
  \def\citenamefont#1{#1}\fi
\expandafter\ifx\csname url\endcsname\relax
  \def\url#1{\texttt{#1}}\fi
\expandafter\ifx\csname urlprefix\endcsname\relax\def\urlprefix{URL }\fi
\providecommand{\bibinfo}[2]{#2}
\providecommand{\eprint}[2][]{\url{#2}}

\bibitem[{\citenamefont{Orton and Foxon}(1998)}]{orton}
\bibinfo{author}{\bibfnamefont{J.~W.} \bibnamefont{Orton}} \bibnamefont{and}
  \bibinfo{author}{\bibfnamefont{C.~T.} \bibnamefont{Foxon}},
  \bibinfo{journal}{Rep. Prog. Phys.} \textbf{\bibinfo{volume}{61}},
  \bibinfo{pages}{1} (\bibinfo{year}{1998}).

\bibitem[{\citenamefont{Widmann et~al.}(1998)\citenamefont{Widmann, Daudin,
  Feuillet, Samson, Rouvi{\`e}re, and Pelekanos}}]{widmann}
\bibinfo{author}{\bibfnamefont{F.}~\bibnamefont{Widmann}},
  \bibinfo{author}{\bibfnamefont{B.}~\bibnamefont{Daudin}},
  \bibinfo{author}{\bibfnamefont{G.}~\bibnamefont{Feuillet}},
  \bibinfo{author}{\bibfnamefont{Y.}~\bibnamefont{Samson}},
  \bibinfo{author}{\bibfnamefont{J.~L.} \bibnamefont{Rouvi{\`e}re}},
  \bibnamefont{and}
  \bibinfo{author}{\bibfnamefont{N.}~\bibnamefont{Pelekanos}},
  \bibinfo{journal}{J. Appl. Phys.} \textbf{\bibinfo{volume}{83}},
  \bibinfo{pages}{7618} (\bibinfo{year}{1998}).

\bibitem[{\citenamefont{Bhaviripudi et~al.}(2007)\citenamefont{Bhaviripudi, Qi,
  Hu, and Belcher}}]{sreekar}
\bibinfo{author}{\bibfnamefont{S.}~\bibnamefont{Bhaviripudi}},
  \bibinfo{author}{\bibfnamefont{J.}~\bibnamefont{Qi}},
  \bibinfo{author}{\bibfnamefont{E.~L.} \bibnamefont{Hu}}, \bibnamefont{and}
  \bibinfo{author}{\bibfnamefont{A.~M.} \bibnamefont{Belcher}},
  \bibinfo{journal}{Nano Lett.} \textbf{\bibinfo{volume}{7}},
  \bibinfo{pages}{3512} (\bibinfo{year}{2007}).

\bibitem[{\citenamefont{Mi{\'c}i{\'c} et~al.}(1999)\citenamefont{Mi{\'c}i{\'c},
  Ahrenkiel, Bertram, and Nozik}}]{micic99}
\bibinfo{author}{\bibfnamefont{O.~I.} \bibnamefont{Mi{\'c}i{\'c}}},
  \bibinfo{author}{\bibfnamefont{S.~P.} \bibnamefont{Ahrenkiel}},
  \bibinfo{author}{\bibfnamefont{D.}~\bibnamefont{Bertram}}, \bibnamefont{and}
  \bibinfo{author}{\bibfnamefont{A.~J.} \bibnamefont{Nozik}},
  \bibinfo{journal}{Appl. Phys. Lett.} \textbf{\bibinfo{volume}{75}},
  \bibinfo{pages}{478} (\bibinfo{year}{1999}).

\bibitem[{\citenamefont{Wang and Herron}(1990)}]{wang42}
\bibinfo{author}{\bibfnamefont{Y.}~\bibnamefont{Wang}} \bibnamefont{and}
  \bibinfo{author}{\bibfnamefont{N.}~\bibnamefont{Herron}},
  \bibinfo{journal}{Phys. Rev. B} \textbf{\bibinfo{volume}{42}},
  \bibinfo{pages}{7253} (\bibinfo{year}{1990}).

\bibitem[{\citenamefont{Wang and Herron}(1991)}]{wang91}
\bibinfo{author}{\bibfnamefont{Y.}~\bibnamefont{Wang}} \bibnamefont{and}
  \bibinfo{author}{\bibfnamefont{N.}~\bibnamefont{Herron}},
  \bibinfo{journal}{J. Phys. Chem.} \textbf{\bibinfo{volume}{95}},
  \bibinfo{pages}{525} (\bibinfo{year}{1991}).

\bibitem[{\citenamefont{Albe et~al.}(1998{\natexlab{b}})\citenamefont{Albe,
  Jouanin, and Bertho}}]{albe57}
\bibinfo{author}{\bibfnamefont{V.}~\bibnamefont{Albe}},
  \bibinfo{author}{\bibfnamefont{C.}~\bibnamefont{Jouanin}}, \bibnamefont{and}
  \bibinfo{author}{\bibfnamefont{D.}~\bibnamefont{Bertho}},
  \bibinfo{journal}{Phys. Rev. B} \textbf{\bibinfo{volume}{57}},
  \bibinfo{pages}{8778} (\bibinfo{year}{1998}{\natexlab{b}}).

\bibitem[{\citenamefont{Albe et~al.}(1998{\natexlab{a}})\citenamefont{Albe,
  Jouanin, and Bertho}}]{albe}
\bibinfo{author}{\bibfnamefont{V.}~\bibnamefont{Albe}},
  \bibinfo{author}{\bibfnamefont{C.}~\bibnamefont{Jouanin}}, \bibnamefont{and}
  \bibinfo{author}{\bibfnamefont{D.}~\bibnamefont{Bertho}},
  \bibinfo{journal}{Phys. Rev. B} \textbf{\bibinfo{volume}{58}},
  \bibinfo{pages}{4713} (\bibinfo{year}{1998}{\natexlab{a}}).


\bibitem[{\citenamefont{P{\'e}rez-Conde and Bhattacharjee}(1999)}]{perez110}
\bibinfo{author}{\bibfnamefont{J.}~\bibnamefont{P{\'e}rez-Conde}}
  \bibnamefont{and} \bibinfo{author}{\bibfnamefont{A.~K.}
  \bibnamefont{Bhattacharjee}}, \bibinfo{journal}{Solid State Commun.}
  \textbf{\bibinfo{volume}{110}}, \bibinfo{pages}{259} (\bibinfo{year}{1999}).

\bibitem[{\citenamefont{Sapra and Sarma}(2004)}]{sapra}
\bibinfo{author}{\bibfnamefont{S.}~\bibnamefont{Sapra}} \bibnamefont{and}
  \bibinfo{author}{\bibfnamefont{D.~D.} \bibnamefont{Sarma}},
  \bibinfo{journal}{Phys. Rev. B} \textbf{\bibinfo{volume}{69}},
  \bibinfo{pages}{125304} (\bibinfo{year}{2004}).

\bibitem[{\citenamefont{Echeverr{\'i}a-Arrondo
  et~al.}(2008)\citenamefont{Echeverr{\'i}a-Arrondo, P{\'e}rez-Conde, and
  Bhattacharjee}}]{carlos-tb}
\bibinfo{author}{\bibfnamefont{C.}~\bibnamefont{Echeverr{\'i}a-Arrondo}},
  \bibinfo{author}{\bibfnamefont{J.}~\bibnamefont{P{\'e}rez-Conde}},
  \bibnamefont{and} \bibinfo{author}{\bibfnamefont{A.~K.}
  \bibnamefont{Bhattacharjee}}, \bibinfo{journal}{J. Appl. Phys.}
  \textbf{\bibinfo{volume}{104}}, \bibinfo{pages}{044308} (\bibinfo{year}{2008}).

\bibitem[{\citenamefont{Biswas et~al.}(2006)\citenamefont{Biswas, Sardar, and
  Rao}}]{biswas}
\bibinfo{author}{\bibfnamefont{K.}~\bibnamefont{Biswas}},
  \bibinfo{author}{\bibfnamefont{K.}~\bibnamefont{Sardar}}, \bibnamefont{and}
  \bibinfo{author}{\bibfnamefont{C.~N.~R.} \bibnamefont{Rao}},
  \bibinfo{journal}{Appl. Phys. Lett.} \textbf{\bibinfo{volume}{89}},
  \bibinfo{pages}{132503} (\bibinfo{year}{2006}).

\bibitem[{\citenamefont{Das et~al.}(2003)\citenamefont{Das, Rao, and
  Jena}}]{das}
\bibinfo{author}{\bibfnamefont{G.~P.} \bibnamefont{Das}},
  \bibinfo{author}{\bibfnamefont{C.~N.~R.} \bibnamefont{Rao}},
  \bibnamefont{and} \bibinfo{author}{\bibfnamefont{P.}~\bibnamefont{Jena}},
  \bibinfo{journal}{Phys. Rev. B} \textbf{\bibinfo{volume}{68}},
  \bibinfo{pages}{035207} (\bibinfo{year}{2003}).



\bibitem[{\citenamefont{d}(2004)\citenamefont{d},
  }]{SIC-2}
\bibinfo{author}{\bibfnamefont{M.} \bibnamefont{Wierzbowska}},
  \bibinfo{author}{\bibfnamefont{D.}~\bibnamefont{S\'anchez-Portal}},
  \bibnamefont{and} \bibinfo{author}{\bibfnamefont{S.}~\bibnamefont{Sanvito}},
  \bibinfo{journal}{Phys. Rev. B} \textbf{\bibinfo{volume}{70}},
  \bibinfo{pages}{235209} (\bibinfo{year}{2004}).






\bibitem[{\citenamefont{Hynninen et~al.}(2006)\citenamefont{Hynninen, Raebiger,
  {von Boehm}, and Ayuela}}]{apl-ayuela}
\bibinfo{author}{\bibfnamefont{T.}~\bibnamefont{Hynninen}},
  \bibinfo{author}{\bibfnamefont{H.}~\bibnamefont{Raebiger}},
  \bibinfo{author}{\bibfnamefont{J.}~\bibnamefont{{von Boehm}}},
  \bibnamefont{and} \bibinfo{author}{\bibfnamefont{A.}~\bibnamefont{Ayuela}},
  \bibinfo{journal}{Appl. Phys. Lett.} \textbf{\bibinfo{volume}{88}},
  \bibinfo{pages}{122501} (\bibinfo{year}{2006}).



\bibitem[{\citenamefont{Hynninen et~al.}(2006)\citenamefont{Hynninen, Raebiger,
  {von Boehm}, and Ayuela}}]{f1}
\bibinfo{author}{\bibfnamefont{S.}~\bibnamefont{Dhar}},
  \bibinfo{author}{\bibfnamefont{O.}~\bibnamefont{Brandt}},
  \bibinfo{author}{\bibfnamefont{A.}~\bibnamefont{{Trampert}}},
\bibinfo{author}{\bibfnamefont{L.}~\bibnamefont{D\"aweritz}},
  \bibinfo{author}{\bibfnamefont{K.~J.}~\bibnamefont{Friedland}},
  \bibinfo{author}{\bibfnamefont{K.~H.}~\bibnamefont{{Ploog}}},
\bibinfo{author}{\bibfnamefont{J.}~\bibnamefont{Keller}},
  \bibinfo{author}{\bibfnamefont{B.}~\bibnamefont{Beshoten}},
  \bibnamefont{and} \bibinfo{author}{\bibfnamefont{G.}~\bibnamefont{G\"untherodt}},
  \bibinfo{journal}{Appl. Phys. Lett.} \textbf{\bibinfo{volume}{82}},
  \bibinfo{pages}{2077} (\bibinfo{year}{2003}).

\bibitem[{\citenamefont{Hynninen et~al.}(2006)\citenamefont{Hynninen, Raebiger,
  {von Boehm}, and Ayuela}}]{f2}
\bibinfo{author}{\bibfnamefont{M.}~\bibnamefont{Zaj\c ac}},
  \bibinfo{author}{\bibfnamefont{J.}~\bibnamefont{Gosk}},
  \bibinfo{author}{\bibfnamefont{G.}~\bibnamefont{{Grazanka}}},
\bibinfo{author}{\bibfnamefont{M.}~\bibnamefont{Kami\'nska}},
  \bibinfo{author}{\bibfnamefont{A.}~\bibnamefont{Twardowski}},
  \bibinfo{author}{\bibfnamefont{B.}~\bibnamefont{{Strojek}}},
\bibinfo{author}{\bibfnamefont{T.}~\bibnamefont{Szyszko}},
  \bibnamefont{and} \bibinfo{author}{\bibfnamefont{S.}~\bibnamefont{Podsiad\l o}},
  \bibinfo{journal}{J. Appl. Phys.} \textbf{\bibinfo{volume}{93}},
  \bibinfo{pages}{4715} (\bibinfo{year}{2003}).









\bibitem[{\citenamefont{Norris et~al.}(2001)\citenamefont{Norris, Yao,
  Charnock, and Kennedy}}]{norris}
\bibinfo{author}{\bibfnamefont{D.~J.} \bibnamefont{Norris}},
  \bibinfo{author}{\bibfnamefont{N.}~\bibnamefont{Yao}},
  \bibinfo{author}{\bibfnamefont{F.~T.} \bibnamefont{Charnock}},
  \bibnamefont{and} \bibinfo{author}{\bibfnamefont{T.~A.}
  \bibnamefont{Kennedy}}, \bibinfo{journal}{Nano Lett.}
  \textbf{\bibinfo{volume}{1}}, \bibinfo{pages}{3} (\bibinfo{year}{2001}).

\bibitem[{\citenamefont{Besombes et~al.}(2004)\citenamefont{Besombes, Leger,
  Maingault, Ferrand, Mariette, and Cibert}}]{besombes}
\bibinfo{author}{\bibfnamefont{L.}~\bibnamefont{Besombes}},
  \bibinfo{author}{\bibfnamefont{Y.}~\bibnamefont{Leger}},
  \bibinfo{author}{\bibfnamefont{L.}~\bibnamefont{Maingault}},
  \bibinfo{author}{\bibfnamefont{D.}~\bibnamefont{Ferrand}},
  \bibinfo{author}{\bibfnamefont{H.}~\bibnamefont{Mariette}}, \bibnamefont{and}
  \bibinfo{author}{\bibfnamefont{J.}~\bibnamefont{Cibert}},
  \bibinfo{journal}{Phys. Rev. Lett.} \textbf{\bibinfo{volume}{93}},
  \bibinfo{pages}{207403} (\bibinfo{year}{2004}).

\bibitem[{\citenamefont{Hofmann et~al.}(2007)\citenamefont{Hofmann, Graf,
  Boeglin, and R{\"u}hl}}]{hofmann2007}
\bibinfo{author}{\bibfnamefont{A.}~\bibnamefont{Hofmann}},
  \bibinfo{author}{\bibfnamefont{C.}~\bibnamefont{Graf}},
  \bibinfo{author}{\bibfnamefont{C.}~\bibnamefont{Boeglin}}, \bibnamefont{and}
  \bibinfo{author}{\bibfnamefont{E.}~\bibnamefont{R{\"u}hl}},
  \bibinfo{journal}{ChemPhysChem.} \textbf{\bibinfo{volume}{8}},
  \bibinfo{pages}{2008} (\bibinfo{year}{2007}).

\bibitem[{\citenamefont{d}(2004)\citenamefont{d}}]{schneider}
\bibinfo{author}{\bibfnamefont{J.}~\bibnamefont{Schneider}},
  \bibinfo{author}{\bibfnamefont{W.}~\bibnamefont{Wilkening}}, 
\bibinfo{author}{\bibfnamefont{M.}~\bibnamefont{Baeumler}}, 
\bibnamefont{and}
  \bibinfo{author}{\bibfnamefont{F.} \bibnamefont{K\"ohl}},
  \bibinfo{journal}{Phys.~Rev.~Lett.} \textbf{\bibinfo{volume}{59}},
  \bibinfo{pages}{240} (\bibinfo{year}{1987}).


\bibitem[{\citenamefont{Raebiger et~al.}(2004)\citenamefont{Raebiger, Ayuela,
  and Nieminen}}]{ayuela1}
\bibinfo{author}{\bibfnamefont{H.}~\bibnamefont{Raebiger}},
  \bibinfo{author}{\bibfnamefont{A.}~\bibnamefont{Ayuela}}, \bibnamefont{and}
  \bibinfo{author}{\bibfnamefont{R.~M.} \bibnamefont{Nieminen}},
  \bibinfo{journal}{J. Phys.: Condens. Matter} \textbf{\bibinfo{volume}{16}},
  \bibinfo{pages}{L457} (\bibinfo{year}{2004}).

\bibitem[{\citenamefont{Raebiger et~al.}(2005)\citenamefont{Raebiger, Ayuela,
  and {von Boehm}}}]{ayuela2}
\bibinfo{author}{\bibfnamefont{H.}~\bibnamefont{Raebiger}},
  \bibinfo{author}{\bibfnamefont{A.}~\bibnamefont{Ayuela}}, \bibnamefont{and}
  \bibinfo{author}{\bibfnamefont{J.}~\bibnamefont{{von Boehm}}},
  \bibinfo{journal}{Phys. Rev. B} \textbf{\bibinfo{volume}{72}},
  \bibinfo{pages}{014465} (\bibinfo{year}{2005}).



\bibitem[{\citenamefont{Marcet et~al.}(2006)\citenamefont{Marcet, Ferrand,
  Kuroda, Gheeraert, Galera, Cibert, and Mariette}}]{marcet}
\bibinfo{author}{\bibfnamefont{S.}~\bibnamefont{Marcet}},
  \bibinfo{author}{\bibfnamefont{D.}~\bibnamefont{Ferrand}},
  \bibinfo{author}{\bibfnamefont{S.}~\bibnamefont{Kuroda}},
  \bibinfo{author}{\bibfnamefont{E.}~\bibnamefont{Gheeraert}},
  \bibinfo{author}{\bibfnamefont{R.~M.} \bibnamefont{Galera}},
  \bibinfo{author}{\bibfnamefont{J.}~\bibnamefont{Cibert}}, \bibnamefont{and}
  \bibinfo{author}{\bibfnamefont{H.}~\bibnamefont{Mariette}},
  \bibinfo{journal}{Mater. Sci. Eng. B} \textbf{\bibinfo{volume}{126}},
  \bibinfo{pages}{240} (\bibinfo{year}{2006}).




\bibitem[{\citenamefont{Moreno et~al.}(2002)\citenamefont{Moreno, Trampert,
  Jenichen, D{\"a}weritz, and Ploog}}]{granular-films}
\bibinfo{author}{\bibfnamefont{M.}~\bibnamefont{Moreno}},
  \bibinfo{author}{\bibfnamefont{A.}~\bibnamefont{Trampert}},
  \bibinfo{author}{\bibfnamefont{B.}~\bibnamefont{Jenichen}},
  \bibinfo{author}{\bibfnamefont{L.}~\bibnamefont{D{\"a}weritz}},
  \bibnamefont{and} \bibinfo{author}{\bibfnamefont{K.~H.} \bibnamefont{Ploog}},
  \bibinfo{journal}{J. Appl. Phys.} \textbf{\bibinfo{volume}{92}},
  \bibinfo{pages}{4672} (\bibinfo{year}{2002}).

\bibitem[{\citenamefont{Janik}(2005)}]{granular-2}
\bibinfo{author}{\bibfnamefont{J.~F.} \bibnamefont{Janik}},
  \bibinfo{journal}{Powder Tech.} \textbf{\bibinfo{volume}{152}},
  \bibinfo{pages}{118} (\bibinfo{year}{2005}).

\bibitem[{\citenamefont{Gosk et~al.}(2006)\citenamefont{Gosk, Dryga{\'s},
  Janik, Palczewska, Paine, and Twardowski}}]{granular}
\bibinfo{author}{\bibfnamefont{J.~B.} \bibnamefont{Gosk}},
  \bibinfo{author}{\bibfnamefont{M.}~\bibnamefont{Dryga{\'s}}},
  \bibinfo{author}{\bibfnamefont{J.~F.} \bibnamefont{Janik}},
  \bibinfo{author}{\bibfnamefont{M.}~\bibnamefont{Palczewska}},
  \bibinfo{author}{\bibfnamefont{R.~T.} \bibnamefont{Paine}}, \bibnamefont{and}
  \bibinfo{author}{\bibfnamefont{A.}~\bibnamefont{Twardowski}},
  \bibinfo{journal}{J. Phys. D: Appl. Phys.} \textbf{\bibinfo{volume}{39}},
  \bibinfo{pages}{3717} (\bibinfo{year}{2006}).


\bibitem[{\citenamefont{Kresse and Hafner}(1993)}]{kresse1}
\bibinfo{author}{\bibfnamefont{G.}~\bibnamefont{Kresse}} \bibnamefont{and}
  \bibinfo{author}{\bibfnamefont{J.}~\bibnamefont{Hafner}},
  \bibinfo{journal}{Phys. Rev. B} \textbf{\bibinfo{volume}{47}},
  \bibinfo{pages}{558} (\bibinfo{year}{1993}).

\bibitem[{\citenamefont{Kresse and Furthm{\"u}ller}(1996)}]{kresse2}
\bibinfo{author}{\bibfnamefont{G.}~\bibnamefont{Kresse}} \bibnamefont{and}
  \bibinfo{author}{\bibfnamefont{J.}~\bibnamefont{Furthm{\"u}ller}},
  \bibinfo{journal}{Phys. Rev. B} \textbf{\bibinfo{volume}{54}},
  \bibinfo{pages}{11169} (\bibinfo{year}{1996}).

\bibitem[{\citenamefont{Kresse and Joubert}(1999)}]{kresse3}
\bibinfo{author}{\bibfnamefont{G.}~\bibnamefont{Kresse}} \bibnamefont{and}
  \bibinfo{author}{\bibfnamefont{D.}~\bibnamefont{Joubert}},
  \bibinfo{journal}{Phys. Rev. B} \textbf{\bibinfo{volume}{59}},
  \bibinfo{pages}{1758} (\bibinfo{year}{1999}).

\bibitem[{\citenamefont{Perdew et~al.}(1996)\citenamefont{Perdew, Burke, and
  Ernzerhof}}]{pbe}
\bibinfo{author}{\bibfnamefont{J.~P.} \bibnamefont{Perdew}},
  \bibinfo{author}{\bibfnamefont{K.}~\bibnamefont{Burke}}, \bibnamefont{and}
  \bibinfo{author}{\bibfnamefont{M.}~\bibnamefont{Ernzerhof}},
  \bibinfo{journal}{Phys. Rev. Lett.} \textbf{\bibinfo{volume}{77}},
  \bibinfo{pages}{3865} (\bibinfo{year}{1996}).

\bibitem[{\citenamefont{Kresse and Joubert}(1999)}]{u1}
\bibinfo{author}{\bibfnamefont{I.~V.}~\bibnamefont{Solovyev}}
\bibnamefont{and}
\bibinfo{author}{\bibfnamefont{P.~H.}~\bibnamefont{Dederichs}},
  \bibinfo{journal}{Phys. Rev. B} \textbf{\bibinfo{volume}{49}},
  \bibinfo{pages}{6736} (\bibinfo{year}{1994}).


\bibitem[{\citenamefont{d}(2003)\citenamefont{d},
  }]{SIC-1}
\bibinfo{author}{\bibfnamefont{B.} \bibnamefont{Sanyal}},
  \bibinfo{author}{\bibfnamefont{O.}~\bibnamefont{Bengone}},
  \bibnamefont{and} \bibinfo{author}{\bibfnamefont{S.}~\bibnamefont{Mirbt}},
  \bibinfo{journal}{Phys. Rev. B} \textbf{\bibinfo{volume}{68}},
  \bibinfo{pages}{205210} (\bibinfo{year}{2003}).

\bibitem[{\citenamefont{Kresse and Joubert}(1999)}]{u3}
\bibinfo{author}{\bibfnamefont{J.}~\bibnamefont{Kang}}
\bibnamefont{and}
\bibinfo{author}{\bibfnamefont{K.~J.}~\bibnamefont{Chang}},
  \bibinfo{journal}{J. Appl. Phys.} \textbf{\bibinfo{volume}{102}},
  \bibinfo{pages}{083910} (\bibinfo{year}{2007}).

\bibitem[{\citenamefont{Kresse and Joubert}(1999)}]{u4}
\bibinfo{author}{\bibfnamefont{L.}~\bibnamefont{Liu}},
\bibinfo{author}{\bibfnamefont{P.~Y.}~\bibnamefont{Yu}},
\bibinfo{author}{\bibfnamefont{Z.}~\bibnamefont{Ma}},
\bibnamefont{and}
\bibinfo{author}{\bibfnamefont{S.~S.}~\bibnamefont{Mao}},
  \bibinfo{journal}{Phys. Rev. Lett.} \textbf{\bibinfo{volume}{100}},
  \bibinfo{pages}{127203} (\bibinfo{year}{2008}).


\bibitem[{\citenamefont{Kresse and Joubert}(1999)}]{ggau}
\bibinfo{author}{\bibfnamefont{J.~A.}~\bibnamefont{Chan}},
\bibinfo{author}{\bibfnamefont{J.~Z.}~\bibnamefont{Liu}},
\bibinfo{author}{\bibfnamefont{H.}~\bibnamefont{Raebiger}},
\bibinfo{author}{\bibfnamefont{S.}~\bibnamefont{Lany}},
 \bibnamefont{and}
  \bibinfo{author}{\bibfnamefont{A.}~\bibnamefont{Zunger}},
  \bibinfo{journal}{Phys. Rev. B} \textbf{\bibinfo{volume}{78}},
  \bibinfo{pages}{184109} (\bibinfo{year}{2008}).






\bibitem[{\citenamefont{d}(2003)\citenamefont{d},
  }]{sic}
\bibinfo{note}{We note that other corrections to GGA such as the self-interaction correction would lead to similar Mn states.\cite{SIC-1,SIC-2,SIC-22,SIC-3}}

\bibitem[{\citenamefont{d}(2004)\citenamefont{d},}]{park}
\bibinfo{author}{\bibfnamefont{J.~H.}~\bibnamefont{Park}},
  \bibinfo{author}{\bibfnamefont{S.~K.}~\bibnamefont{Kwon}},
  \bibnamefont{and} \bibinfo{author}{\bibfnamefont{B.~I.}~\bibnamefont{Min}},
  \bibinfo{journal}{Physica B} \textbf{\bibinfo{volume}{281\&282}},
  \bibinfo{pages}{703} (\bibinfo{year}{2000}).



\bibitem[{\citenamefont{Huang et~al.}(2005)\citenamefont{Huang, Lindgren, and
  Chelikowsky}}]{chelikowsky}
\bibinfo{author}{\bibfnamefont{X.}~\bibnamefont{Huang}},
  \bibinfo{author}{\bibfnamefont{E.}~\bibnamefont{Lindgren}}, \bibnamefont{and}
  \bibinfo{author}{\bibfnamefont{J.~R.} \bibnamefont{Chelikowsky}},
  \bibinfo{journal}{Phys. Rev. B} \textbf{\bibinfo{volume}{71}},
  \bibinfo{pages}{165328} (\bibinfo{year}{2005}).



\bibitem[{\citenamefont{Huang et~al.}(2005)\citenamefont{-}}]{scell-1}
\bibinfo{author}{\bibfnamefont{Y.}~\bibnamefont{Zhao}},
\bibinfo{author}{\bibfnamefont{Y.-H.}~\bibnamefont{Kim}},
\bibinfo{author}{\bibfnamefont{M.-H.}~\bibnamefont{Du}}, \bibnamefont{and}
  \bibinfo{author}{\bibfnamefont{S.~B.} \bibnamefont{Zhang}},
  \bibinfo{journal}{Phys. Rev. Lett.} \textbf{\bibinfo{volume}{93}},
  \bibinfo{pages}{015502} (\bibinfo{year}{2004}).

\bibitem[{\citenamefont{Huang et~al.}(2005)\citenamefont{-}}]{scell-2}
\bibinfo{author}{\bibfnamefont{M.~C.} \bibnamefont{Qian}},
\bibinfo{author}{\bibfnamefont{C.~Y.} \bibnamefont{Fong}},
\bibinfo{author}{\bibfnamefont{W.~E.} \bibnamefont{Pickett}}, \bibnamefont{and}
  \bibinfo{author}{\bibfnamefont{H.-Y.} \bibnamefont{Wang}},
  \bibinfo{journal}{J. Appl. Phys.} \textbf{\bibinfo{volume}{95}},
  \bibinfo{pages}{7459} (\bibinfo{year}{2004}).


\bibitem[{\citenamefont{Huang et~al.}(2005)\citenamefont{-}}]{scell-3}
\bibinfo{author}{\bibfnamefont{S.~K.}~\bibnamefont{Bhattacharya}} \bibnamefont{and}
  \bibinfo{author}{\bibfnamefont{A.} \bibnamefont{Kshirsagar}},
  \bibinfo{journal}{Eur. Phys. J. D} \textbf{\bibinfo{volume}{48}},
  \bibinfo{pages}{355} (\bibinfo{year}{2008}).

\bibitem[{\citenamefont{Huang et~al.}(2005)\citenamefont{-}}]{scell-4}
\bibinfo{author}{\bibfnamefont{C.}~\bibnamefont{Echeverr\'ia-Arrondo}},
\bibinfo{author}{\bibfnamefont{J.}~\bibnamefont{P\'erez-Conde}}, \bibnamefont{and}
  \bibinfo{author}{\bibfnamefont{A.} \bibnamefont{Ayuela}},
  \bibinfo{journal}{Appl. Phys. Lett.} \textbf{\bibinfo{volume}{95}},
  \bibinfo{pages}{043111} (\bibinfo{year}{2009}).

\bibitem[{\citenamefont{Fatah et~al.}(1994)\citenamefont{Fatah, Piorek,
  Harrison, Stirner, and Hagston}}]{fatah}
\bibinfo{author}{\bibfnamefont{J.~M.} \bibnamefont{Fatah}},
  \bibinfo{author}{\bibfnamefont{J.}~\bibnamefont{Piorek}},
  \bibinfo{author}{\bibfnamefont{P.}~\bibnamefont{Harrison}},
  \bibinfo{author}{\bibfnamefont{T.}~\bibnamefont{Stirner}}, \bibnamefont{and}
  \bibinfo{author}{\bibfnamefont{W.~E.} \bibnamefont{Hagston}},
  \bibinfo{journal}{Phys. Rev. B} \textbf{\bibinfo{volume}{49}},
  \bibinfo{pages}{10341} (\bibinfo{year}{1994}).

\bibitem[{\citenamefont{Echeverr{\'i}a-Arrondo
  et~al.}(2009)\citenamefont{Echeverr{\'i}a-Arrondo, P{\'e}rez-Conde, and
  Ayuela}}]{carlos}
\bibinfo{author}{\bibfnamefont{C.}~\bibnamefont{Echeverr{\'i}a-Arrondo}},
  \bibinfo{author}{\bibfnamefont{J.}~\bibnamefont{P{\'e}rez-Conde}},
  \bibnamefont{and} \bibinfo{author}{\bibfnamefont{A.}~\bibnamefont{Ayuela}},
  \bibinfo{journal}{Phys. Rev. B} \textbf{\bibinfo{volume}{79}},
  \bibinfo{pages}{155319} (\bibinfo{year}{2009}).

\bibitem[{\citenamefont{Schmidt et~al.}(2006)\citenamefont{Schmidt, Venezuela,
  Arantes, and Fazzio}}]{nanowire}
\bibinfo{author}{\bibfnamefont{T.~M.} \bibnamefont{Schmidt}},
  \bibinfo{author}{\bibfnamefont{P.}~\bibnamefont{Venezuela}},
  \bibinfo{author}{\bibfnamefont{J.~T.} \bibnamefont{Arantes}},
  \bibnamefont{and} \bibinfo{author}{\bibfnamefont{A.}~\bibnamefont{Fazzio}},
  \bibinfo{journal}{Phys. Rev. B} \textbf{\bibinfo{volume}{73}},
  \bibinfo{pages}{235330} (\bibinfo{year}{2006}).



%



\bibitem[{\citenamefont{Larson et~al.}(1988)\citenamefont{Larson, Carlsson, and
  Ehrenreich}}]{larson}
\bibinfo{author}{\bibfnamefont{B.~E.} \bibnamefont{Larson}},
  \bibinfo{author}{\bibfnamefont{H.~K.} \bibnamefont{Carlsson}},
  \bibnamefont{and}
  \bibinfo{author}{\bibfnamefont{H.}~\bibnamefont{Ehrenreich}},
  \bibinfo{journal}{Phys. Rev. B} \textbf{\bibinfo{volume}{37}},
  \bibinfo{pages}{4137} (\bibinfo{year}{1988}).

\bibitem[{\citenamefont{d}(1979)\citenamefont{d}}]{gaj}
\bibinfo{author}{\bibfnamefont{J.~A.} \bibnamefont{Gaj}},
  \bibinfo{author}{\bibfnamefont{R.}~\bibnamefont{Planel}}, \bibnamefont{and}
  \bibinfo{author}{\bibfnamefont{G.}~\bibnamefont{Fishman}},
  \bibinfo{journal}{Solid State Commun.} \textbf{\bibinfo{volume}{29}},
  \bibinfo{pages}{435} (\bibinfo{year}{1979}).

\bibitem[{\citenamefont{d}(1983)\citenamefont{d}}]{bhatta}
\bibinfo{author}{\bibfnamefont{A.~K.} \bibnamefont{Bhattacharjee}},
  \bibinfo{journal}{Phys. Rev. B} \textbf{\bibinfo{volume}{58}},
  \bibinfo{pages}{15660} (\bibinfo{year}{1998}).



\bibitem[{\citenamefont{Norberg and Gamelin}(2006)}]{gamelin}
\bibinfo{author}{\bibfnamefont{N.~S.} \bibnamefont{Norberg}} \bibnamefont{and}
  \bibinfo{author}{\bibfnamefont{D.~R.} \bibnamefont{Gamelin}},
  \bibinfo{journal}{J. Appl. Phys.} \textbf{\bibinfo{volume}{99}},
  \bibinfo{pages}{08M104} (\bibinfo{year}{2006}).

\bibitem[{\citenamefont{m}(2007)}]{santangelo}
\bibinfo{author}{\bibfnamefont{P.~L.} \bibnamefont{Archer}},
\bibinfo{author}{\bibfnamefont{S.~A.} \bibnamefont{Santangelo}},
 \bibnamefont{and}
  \bibinfo{author}{\bibfnamefont{D.~R.} \bibnamefont{Gamelin}},
  \bibinfo{journal}{Nano Lett.} \textbf{\bibinfo{volume}{7}},
  \bibinfo{pages}{1037} (\bibinfo{year}{2007}).









\bibitem[{\citenamefont{Bhattacharjee et~al.}(1983)\citenamefont{Bhattacharjee,
  Fishman, and Coqblin}}]{bhatta83}
\bibinfo{author}{\bibfnamefont{A.~K.} \bibnamefont{Bhattacharjee}},
  \bibinfo{author}{\bibfnamefont{G.}~\bibnamefont{Fishman}}, \bibnamefont{and}
  \bibinfo{author}{\bibfnamefont{B.}~\bibnamefont{Coqblin}},
  \bibinfo{journal}{Physica B \& C} \textbf{\bibinfo{volume}{117-118}},
  \bibinfo{pages}{449} (\bibinfo{year}{1983}).



\bibitem[{\citenamefont{Merad et~al.}(2006)\citenamefont{Merad, Kanoun, and
  Goumri-Said}}]{merad}
\bibinfo{author}{\bibfnamefont{A.~E.} \bibnamefont{Merad}},
  \bibinfo{author}{\bibfnamefont{M.~B.} \bibnamefont{Kanoun}},
  \bibnamefont{and}
  \bibinfo{author}{\bibfnamefont{S.}~\bibnamefont{Goumri-Said}},
  \bibinfo{journal}{J. Magn. and Magn. Mater.} \textbf{\bibinfo{volume}{302}},
  \bibinfo{pages}{536} (\bibinfo{year}{2006}).

\bibitem[{\citenamefont{Bhattacharjee and P{\'e}rez-Conde}(2003)}]{bhatta3}
\bibinfo{author}{\bibfnamefont{A.~K.} \bibnamefont{Bhattacharjee}}
  \bibnamefont{and}
  \bibinfo{author}{\bibfnamefont{J.}~\bibnamefont{P{\'e}rez-Conde}},
  \bibinfo{journal}{Phys. Rev. B} \textbf{\bibinfo{volume}{68}},
  \bibinfo{pages}{045303} (\bibinfo{year}{2003}).

\bibitem[{\citenamefont{d}(1987)}]{chanier}
\bibinfo{author}{\bibfnamefont{T.} \bibnamefont{Chanier}},
\bibinfo{author}{\bibfnamefont{M.} \bibnamefont{Sargolzaei}},
\bibinfo{author}{\bibfnamefont{I.} \bibnamefont{Opahle}},
\bibinfo{author}{\bibfnamefont{R.} \bibnamefont{Hayn}},
  \bibnamefont{and}
  \bibinfo{author}{\bibfnamefont{K.}~\bibnamefont{Koepernik}},
  \bibinfo{journal}{Phys. Rev. B} \textbf{\bibinfo{volume}{73}},
  \bibinfo{pages}{134418} (\bibinfo{year}{2006}).



\bibitem[{\citenamefont{d}(1987)}]{shapira}
\bibinfo{author}{\bibfnamefont{Y.} \bibnamefont{Shapira}}
  \bibnamefont{and}
  \bibinfo{author}{\bibfnamefont{N.~F.}~\bibnamefont{Olivera, Jr.}},
  \bibinfo{journal}{Phys. Rev. B} \textbf{\bibinfo{volume}{35}},
  \bibinfo{pages}{6888} (\bibinfo{year}{1987}).

\bibitem[{\citenamefont{d}(1987)}]{stowell}
\bibinfo{author}{\bibfnamefont{C.~A.} \bibnamefont{Stowell}},
\bibinfo{author}{\bibfnamefont{R.~J.} \bibnamefont{Wiacek}},
\bibinfo{author}{\bibfnamefont{A.~E.} \bibnamefont{Saunders}},
  \bibnamefont{and}
  \bibinfo{author}{\bibfnamefont{B.~A.} \bibnamefont{Korgel}},
  \bibinfo{journal}{Nano Lett.} \textbf{\bibinfo{volume}{3}},
  \bibinfo{pages}{1441} (\bibinfo{year}{2003}).





\bibitem[{\citenamefont{d}(1987)}]{sati}
\bibinfo{author}{\bibfnamefont{P.} \bibnamefont{Sati}},
\bibinfo{author}{\bibfnamefont{C.} \bibnamefont{Depais}},
\bibinfo{author}{\bibfnamefont{C.} \bibnamefont{Morhain}},
\bibinfo{author}{\bibfnamefont{S.} \bibnamefont{Sch\"afer}},
  \bibnamefont{and}
  \bibinfo{author}{\bibfnamefont{A.}~\bibnamefont{Stepanov}},
  \bibinfo{journal}{Phys. Rev. Lett.} \textbf{\bibinfo{volume}{98}},
  \bibinfo{pages}{137204} (\bibinfo{year}{2007}).

\bibitem[{\citenamefont{d}(1987)}]{white}
\bibinfo{author}{\bibfnamefont{M.~A.} \bibnamefont{White}},
\bibinfo{author}{\bibfnamefont{S.~T.} \bibnamefont{Ochsenbein}},
  \bibnamefont{and}
  \bibinfo{author}{\bibfnamefont{D.~R.}~\bibnamefont{Gamelin}},
  \bibinfo{journal}{Chem. Mat.} \textbf{\bibinfo{volume}{20}},
  \bibinfo{pages}{7107} (\bibinfo{year}{2008}).




\bibitem[{\citenamefont{d}(1987)}]{couto}
\bibinfo{author}{\bibfnamefont{O.~D.~D.} \bibnamefont{Couto Jr.}},
\bibinfo{author}{\bibfnamefont{M.~J.~S.~P.} \bibnamefont{Brasil}},
\bibinfo{author}{\bibfnamefont{F.} \bibnamefont{Likawa}},
\bibinfo{author}{\bibfnamefont{C.} \bibnamefont{Giles}},
\bibinfo{author}{\bibfnamefont{C.} \bibnamefont{Adriano}},
\bibinfo{author}{\bibfnamefont{J.~R.~R.} \bibnamefont{Bortoleto}},
\bibinfo{author}{\bibfnamefont{M.~A.~A.} \bibnamefont{Pudenzi}},
\bibinfo{author}{\bibfnamefont{H.~R.} \bibnamefont{Gutierrez}},
  \bibnamefont{and}
  \bibinfo{author}{\bibfnamefont{I.}~\bibnamefont{Danilov}},
  \bibinfo{journal}{Appl. Phys. Lett.} \textbf{\bibinfo{volume}{86}},
  \bibinfo{pages}{71906} (\bibinfo{year}{2005}).

\bibitem[{\citenamefont{d}(1987)}]{janik}
\bibinfo{author}{\bibfnamefont{J.} \bibnamefont{Sadowski}},
\bibinfo{author}{\bibfnamefont{E.} \bibnamefont{Janik}},
\bibinfo{author}{\bibfnamefont{E.} \bibnamefont{Lusakowska}},
\bibinfo{author}{\bibfnamefont{J.~Z.} \bibnamefont{Domagala}},
\bibinfo{author}{\bibfnamefont{S.} \bibnamefont{Kret}},
\bibinfo{author}{\bibfnamefont{P.} \bibnamefont{Dlu\.zewski}},
\bibinfo{author}{\bibfnamefont{M.} \bibnamefont{Adell}},
\bibinfo{author}{\bibfnamefont{J.} \bibnamefont{Kanski}},
\bibinfo{author}{\bibfnamefont{L.} \bibnamefont{Ilver}},
\bibinfo{author}{\bibfnamefont{R.} \bibnamefont{Brucas}},
  \bibnamefont{and}
  \bibinfo{author}{\bibfnamefont{M.}~\bibnamefont{Hanson}},
  \bibinfo{journal}{Appl. Phys. Lett.} \textbf{\bibinfo{volume}{87}},
  \bibinfo{pages}{263114} (\bibinfo{year}{2005}).


\bibitem[{\citenamefont{d}(1987)}]{wasik}
\bibinfo{author}{\bibfnamefont{A.} \bibnamefont{Kwiatkowski}},
\bibinfo{author}{\bibfnamefont{D.} \bibnamefont{Wasik}},
\bibinfo{author}{\bibfnamefont{M.} \bibnamefont{Kami\'nska}},
\bibinfo{author}{\bibfnamefont{J.} \bibnamefont{Borysiuk}},
\bibinfo{author}{\bibfnamefont{R.} \bibnamefont{Bo\.zek}},
\bibinfo{author}{\bibfnamefont{J.} \bibnamefont{Sadowski}},
  \bibnamefont{and}
  \bibinfo{author}{\bibfnamefont{A.}~\bibnamefont{Twardowski}},
  \bibinfo{journal}{J. Mater. Sci.: Mater. Electron.} \textbf{\bibinfo{volume}{19}},
  \bibinfo{pages}{740} (\bibinfo{year}{2008}).



\bibitem[{\citenamefont{d}(2006)\citenamefont{d},
  }]{SIC-22}
\bibinfo{author}{\bibfnamefont{A.} \bibnamefont{Filippetti}},
  \bibinfo{author}{\bibfnamefont{N.~A.} \bibnamefont{Spaldin}},
  \bibnamefont{and}
  \bibinfo{author}{\bibfnamefont{S.} \bibnamefont{Sanvito}}
  \bibinfo{journal}{Chem. Phys.} \textbf{\bibinfo{volume}{309}},
  \bibinfo{pages}{59} (\bibinfo{year}{2005}).





\bibitem[{\citenamefont{d}(2006)\citenamefont{d},
  }]{SIC-3}
\bibinfo{author}{\bibfnamefont{M.} \bibnamefont{Toyoda}},
  \bibinfo{author}{\bibfnamefont{H.}~\bibnamefont{Akai}},
  \bibinfo{author}{\bibfnamefont{K.}~\bibnamefont{Sato}},
  \bibnamefont{and} \bibinfo{author}{\bibfnamefont{H.}~\bibnamefont{Katayama-Yoshida}},
  \bibinfo{journal}{Phys. Stat. Sol. C} \textbf{\bibinfo{volume}{3}},
  \bibinfo{pages}{4155} (\bibinfo{year}{2006}).


\end{thebibliography}
\end{document}